\documentclass[twocolumn,showpacs,amsmath,amssymb,aps,prl]{revtex4-1}

\usepackage{graphicx}% Include figure files
\usepackage{dcolumn}% Align table columns on decimal point
\usepackage{bm}% bold math
\usepackage{placeins}
\usepackage[english=american]{csquotes}

\begin{document}

\title{Deviations from Boltzmann-Gibbs equilibrium in confined optical lattices}

\author{Andreas Dechant}
\affiliation{Department of Physics, Bar Ilan University, Ramat-Gan 52900, Israel}
\author{David A. Kessler}
\affiliation{Department of Physics, Bar Ilan University, Ramat-Gan 52900, Israel}
\author{Eli Barkai}
\affiliation{Department of Physics, Bar Ilan University, Ramat-Gan 52900, Israel}

\begin{abstract}
Cold atoms in dissipative optical lattices have long been known to exhibit anomalous kinetics due to an effective nonlinear friction force.
Here we show that confining the spatial motion of the atoms will lead to an anomalous non-Boltzmann-Gibbs equilibrium state, with a power law tail at large energies.
Only in the limit of deep optical lattices, do we regain the Boltzmann-Gibbs state.
For strong confinement relative to the damping, we find an explicit expression for the equilibrium phase-space distribution, which generally differs from the canonical Boltzmann-Gibbs state at all energies.
Both in the low and high energy limits, the equilibrium distribution is a function of the system's Hamiltonian.
At intermediate energies, however, the distribution is not a function of energy only and equipartition is violated.
\end{abstract}


\maketitle

Sisyphus cooling in optical lattices \cite{coh90,cas90,gry01} has been pivotal in cooling neutral atoms to temperatures in the micro-Kelvin range below the Doppler limit and as a stepping stone to reach even lower temperatures and explore new quantum states of matter.
From the point of view of cooling, the Sisyphus mechanism has been thoroughly investigated, in experiments as well as in theory.
However, the nonlinear momentum-dependence of the cooling mechanism, discussed in more detail below, also induces a wealth of unusual and interesting statistical effects, for a recent review see Ref.~\cite{lut13}. 
In particular, the velocity distribution of the cold atoms has power-law tails \cite{lut03,dou06}, contrary to the Gaussian statistics found in ordinary Boltzmann-Gibbs (BG) statistical mechanics. 
From a dynamical point of view, the system exhibits anomalous superdiffusion \cite{mar96,kat97,kes10,bar14,dec12,sag12} for shallow lattices.
The insight that anomalous statistics describes a broad range of systems in physics \cite{bou90,bar02,met04} warrants a second, closer look at cold atoms as an experimental toolbox, not only for quantum mechanics, but also to explore and go beyond the boundaries of traditional statistical mechanics.
Further, in laser cooling, the light field replaces the classical heat bath of temperature $T$, and this raises the question of what replaces the canonical BG distribution and whether the statistical description changes in a fundamental way.
In this letter, we investigate a semiclassical model of atoms in an optical lattice with an additional confining field and discuss the resulting anomalous equilibrium phase-space probability distribution function (PSPDF).
This equilibrium state differs from the BG one, in an intricate but fundamental manner depending on the depth of the lattice and the strength of the confinement.

The BG distribution describes an ensemble of particles with kinetic energy $E_k = p^2/(2 m)$ in contact with an ideal heat bath at temperature $T$ and in a conservative potential $U(x)$, and is given by,
\begin{align}
P_\text{BG}(x,p) = Z^{-1} e^{-\frac{p^2}{2 m k_B T}-\frac{U(x)}{k_B T}} = Z^{-1} e^{-\frac{H(x,p)}{k_B T}} , \label{boltzmann-potential}
\end{align}
with the normalizing partition function $Z$.
An important feature of the BG distribution is that it is a function of the system's Hamiltonian, $H(x,p) = p^2/(2 m) + U(x)$.
This implies that surfaces of constant energy in phase-space are equivalent to surfaces of constant probability.
Equivalently, in terms of ensemble theory, the probability of finding the system in a state with a certain energy depends only on the energy.
An immediate consequence is the equipartition theorem, which, for the paradigmatic harmonic potential $U(x) = m \omega^2 x^2/2$, states that, in equilibrium, the average potential $E_p = U(x)$ energy and kinetic energy $E_k$ are equal and proportional to the temperature, $\langle E_k \rangle = \langle E_p \rangle = \frac{k_B T}{2}$,
where $\langle \ldots \rangle$ denotes the ensemble average with respect to the equilibrium distribution.
The BG distribution \eqref{boltzmann-potential} is the exponential of the Hamiltonian, and thus additive contributions to the system's energy will factorize in terms of the phase-space distribution. 
This allows us to separate the kinetic from the potential degrees of freedom, $P_\text{BG}(x,p) = P_p(p) P_x(x)$.

For atoms under Sisyphus cooling, the situation is more involved.
Within the semiclassical description, the atoms are subject to a nonlinear, momentum-dependent friction force, which encapsulates the cooling mechanism \cite{cas90}.
In addition, there are random recoil events due to spontaneous emission of photons, modeled as Gaussian white noise $\langle \eta(t) \eta(t') \rangle = 2 D_p \delta(t-t')$.
The atoms' dynamics is described by the Langevin equation \cite{cas90,mar96},
\begin{align}
\dot{x}(t) &= \frac{p(t)}{m} , \nonumber \\
\dot{p}(t) &= \underbrace{- \gamma \frac{p(t)}{1 + p^2(t)/p_c^2}}_\text{friction force} \; \underbrace{\vphantom{ \frac{\gamma p}{1 + p^2/p_c^2} }-  U'(x(t))}_\text{confinement} \; \underbrace{\vphantom{ \frac{\gamma p}{1 + p^2/p_c^2} }+ \eta(t)}_\text{noise} \label{langevin},
\end{align}
where $\gamma$ is the damping coefficient, $p_c$ is the capture momentum and $D_p$ the momentum diffusion coefficient \footnote{Generally, the diffusion coefficient coefficient depends on momentum, $D_p(p)$ \cite{cas90}. However, this dependence does not change the qualitative nature of the results \cite{mar96}.}, all of which can be expressed in terms of the parameters of the optical lattice.
At small momenta $|p| \ll p_c$, the friction force is Stokes-like, $F_f(p) \propto - p$; i.e.~linear in the momentum.
If the friction was linear at all momenta, we would obtain precisely the BG distribution \eqref{boltzmann-potential} with $k_B T = D_p/(\gamma m)$ for the equilibrium state of the system.
However, at large momenta $|p| \gg p_c$ the magnitude of the friction force decreases as $F_f(p) \propto - 1/p$ with increasing momentum -- the cooling mechanism fails for very fast atoms.
This nonlinearity of the friction force induces temporal correlations in the motion of the atoms \cite{mar96}, as fast atoms experience only a weak friction force and thus tend to stay fast.
These correlations are responsible for the power-law statistics and anomalous dynamics \cite{dec12a,bar14}.
These have been discussed in detail in previous work \cite{mar96,lut03,kes10,dec12a} for the case without a confining potential $U(x)$.
There the atomic cloud spreads \mbox{(super-)} diffusively \cite{sag12,kes12,dec12,bar14} and thus there is no stationary PSPDF.
What happens if we confine the atoms spatially?
If the system obeyed BG statistics, the joint distribution of momentum and position would factorize, so that the momentum distribution would be independent of the confining field $U(x)$ and equipartition would hold.
However, as we will discuss in the following, neither is generally the case.
We will focus here on the case of a harmonic potential $U(x) = m \omega^2 x^2/2$.
The PSPDF of the ensemble of random motions described by Eq.~\eqref{langevin} is given by the Klein-Kramers equation \cite{ris86}.
For convenience of notation, we switch to dimensionless variables $m \omega x/p_c \rightarrow x$, $p/p_c \rightarrow p$, 
\begin{align}
\bigg[\Omega \Big[ x \partial_p - p \partial_x \Big] + \partial_p \Big[ \frac{p}{1+p^2} + D \partial_p \Big] - \partial_t \bigg] P(x,p,t) = 0 . \label{klein-kramers}
\end{align}
We here defined the dimensionless parameters $\Omega = \omega/\gamma$, which quantifies the strength of the confining potential, and $D = D_p/(\gamma p_c^2)$, which is related to the depth of the optical lattice $U_0$ by $D = c E_r/U_0$ with the photon recoil energy $E_r$ and a constant $c \sim \mathcal{O}(10)$ that depends on the precise details of the experimental system \cite{mar96,kes10}.

We discuss the properties of the stationary solution, $\partial_t P(x,p,t) = 0$, of Eq.~\eqref{klein-kramers} and how it compares to the BG distribution.
We phrase this in terms of two questions: Whether and under what circumstances the PSPDF is solely a function of the Hamiltonian and, if so, whether this function is exponential.
Note that in the following, we always consider the case $D < 1$, since only here a stationary state exists, the case $D > 1$ will be discussed elsewhere \cite{upcoming}.

\begin{figure}[h!]
\includegraphics[width=0.48\textwidth, clip, trim=0cm 0cm 0cm 0cm]{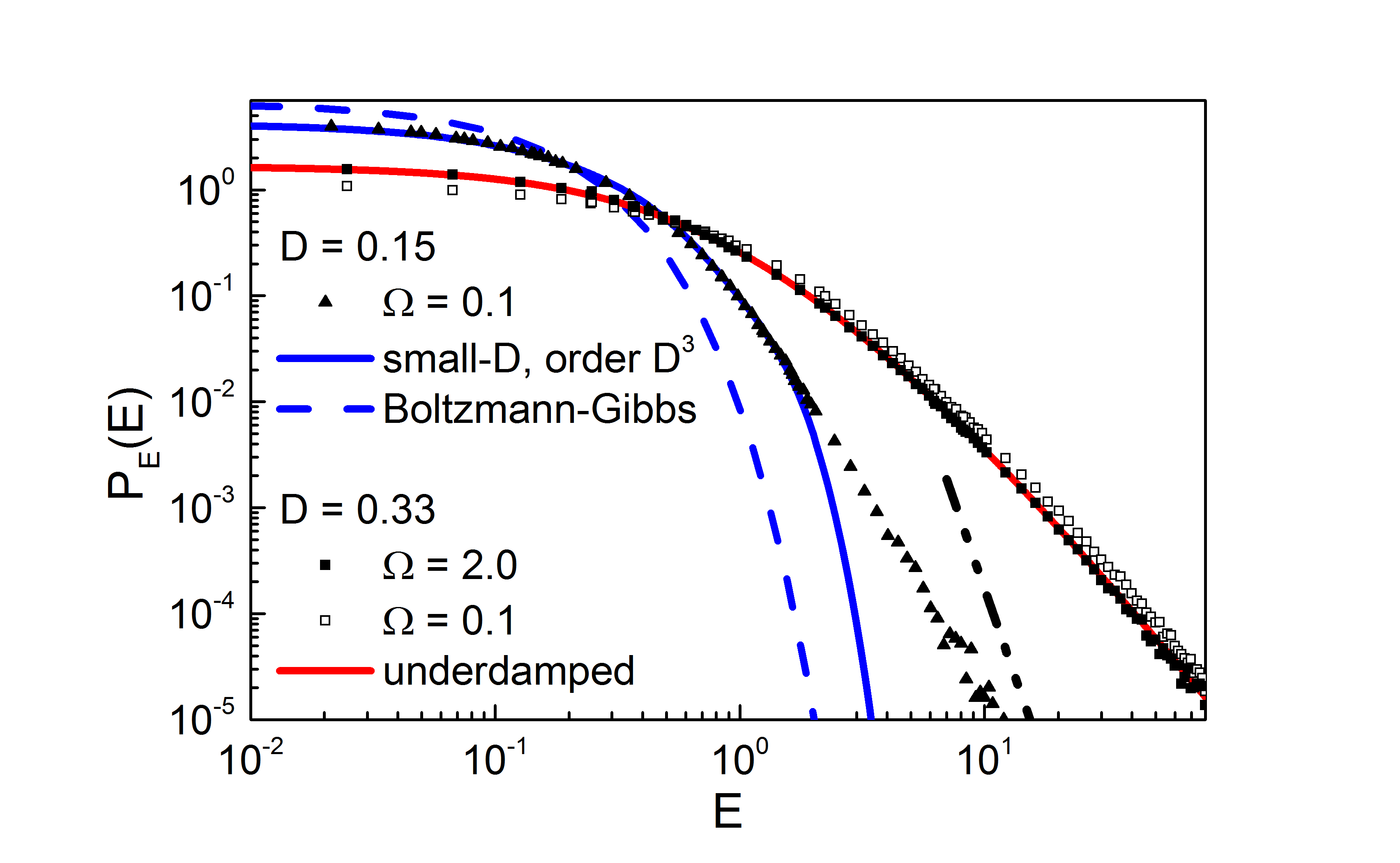}
\caption{Energy PDFs for $D=1/3$ and $D=3/20$. Symbols are data from numerical simulations of Eq.~\eqref{langevin}, colored lines are analytical results. For the larger value of $D = 1/3$ (red, squares), the underdamped approximation Eq.~\eqref{stationary-density} is accurate even for moderate frequencies $\Omega = 2.0$ (full squares). For small frequencies $\Omega = 0.1$ (empty squares), the energy PDF differs in the center, the tails, however, still follow the same power-law behavior $P_E(E) \propto E^{-1/D}$. For smaller values of $D$ (blue, triangles), the PDF has more weight in the center and is there described by the small-$D$ expansion Eq.~\eqref{small-energy} (full line), which considerably improves upon the naive BG distribution Eq.~\eqref{boltzmann-potential} (dashed line). The the large energy power-law (dash-dotted line) is approached very slowly.}
\label{fig:energy-dist}
\end{figure}

\begin{figure}[h!]
\includegraphics[width=0.48\textwidth, clip, trim=0cm 0cm 0cm 0cm]{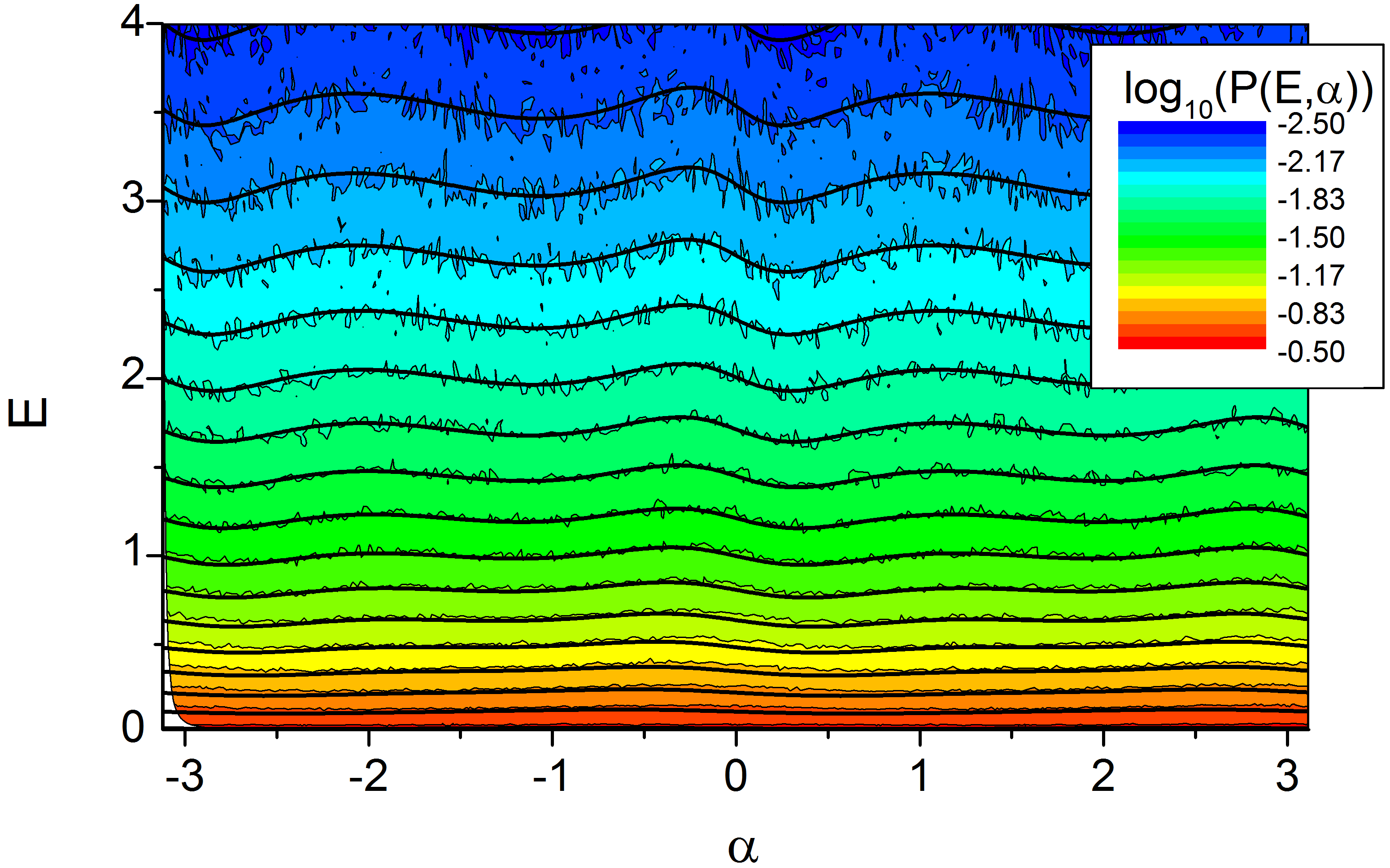}
\caption{PSPDF for $D = 1/3$ and $\Omega = 2$ as a function of energy $E$ and angle $\alpha$. The colored areas are the result of numerical Langevin simulations, the bold contours are the analytical results from the underdamped approximation \eqref{underdamped-expansion} up to order $\Omega^{-1}$. While for small energies, the equi-probability lines are approximately straight lines, for intermediate energies, the angle-dependence is clearly visible.}
\label{fig:underdamped-joint-dist}
\end{figure}

\textbf{Strong confinement $\Omega \gg 1$.} 
We expect the system to be accurately described by its energy whenever its evolution is approximately Hamiltonian; i.e., when dissipation and noise can be treated as small perturbations.
More precisely, this so-called underdamped limit \cite{kra40,str63,han90} is defined by the condition that the change in energy over one period of the Hamiltonian evolution is small compared to the total energy.
We can formalize this by introducing the energy $E = (p^2+x^2)/2$ and the phase-space angle $\alpha = \arctan(p/x)$ in terms of which the stationary limit of Eq.~\eqref{klein-kramers} reads
\begin{align}
\bigg[\partial_{\alpha} + &\Omega^{-1} \mathcal{L}_E \bigg] P(E,\alpha) = 0 \qquad \text{with} \label{klein-kramers-energy} \\
\mathcal{L}_E &= \partial_\alpha \frac{\sin(\alpha)\cos(\alpha)}{1+2 E \sin^2(\alpha)} + \partial_E \frac{2 E \sin^2(\alpha)}{1+2 E \sin^2(\alpha)}  \nonumber \\
& + D \bigg[ 2 \partial_\alpha \sin(\alpha) \cos(\alpha) \partial_E  + \frac{1}{2 E} \partial_\alpha \cos^2(\alpha) \partial_\alpha \nonumber \\
& + (\sin^2(\alpha) - \cos^2(\alpha)) \partial_E + 2 \sin^2(\alpha) \partial_E E \partial_E \bigg] . \nonumber
\end{align}
Here the operator $\mathcal{L}_E$ contains the terms due to friction and noise, while the operator $\partial_{\alpha}$ describes the Hamiltonian part of the dynamics.
Obviously, the underdamped approximation holds for $\Omega \gg 1$, where we have to leading order $\partial_\alpha P(E,\alpha) = 0$ and thus,
\begin{align}
%\partial_\alpha P(E,\alpha) &\simeq \mathcal{O}(\Omega^{-1}) \nonumber \\
P(E,\alpha) &\simeq P_E(E)/(2 \pi) + \mathcal{O}(\Omega^{-1}) . \label{underdamped-leading}
\end{align}
Plugging this into Eq.~\eqref{klein-kramers-energy} and integrating over $\alpha$, we find an equation for the stationary energy PDF,
\begin{align}
\bigg[ \partial_E \Big( 1 - \frac{1}{\sqrt{1+2E}} \Big) + D \partial_E E \partial_E \bigg] P_E(E) = 0 . \label{energy-diffusion-stationary}
\end{align}
The normalized solution to this stationary energy-diffusion equation reads, for $D < 1$,
\begin{align}
P_E(E) = \frac{2^{\frac{2}{D}} (1-D)(2-D)}{2 D} \Big(1 + \sqrt{1+2E}\Big)^{-\frac{2}{D}} \label{stationary-density}. 
\end{align}
Changing back to position and momentum, we obtain to leading order a PSPDF $P(x,p) = P_E(H(x,p))/(2 \pi)$ that depends only on the Hamiltonian.
Contrary to the BG density, however, the distribution is not exponential in the Hamiltonian, and does not factorize into a potential and kinetic part.
The energy PDF \eqref{stationary-density} is compared to the results from numerical simulations in Fig.~\ref{fig:energy-dist} and shows excellent agreement with the latter already for moderate values of the frequency.
Asymptotically for large energies, the energy PDF Eq.~\eqref{stationary-density} behaves as a power law $P_E(E) \propto E^{-1/D}$.
Intriguingly, Eq.~\eqref{stationary-density} tends to the BG distribution in the limit $D \rightarrow 0$, which is reminiscent of a $q$-exponential and Tsallis statistics \cite{tsa88,lut03}.
The momentum PDF decays as $P_p(p) \propto p^{-2/D+1}$ for large momenta, which is markedly different from the exponent $P_p(p) \propto p^{-1/D}$ obtained without the confinement \cite{cas90}.
The power-law form of the energy PDF immediately implies that the average energy,
\begin{align}
\langle E \rangle = \frac{D(2-D)}{(2-3 D)(1-2 D)}, \label{average-energy}
\end{align}
diverges in the stationary state for $D > 1/2$.
Importantly, the average kinetic energy $\langle E_k \rangle = \langle E \rangle / 2$ is always smaller than the result found for the unconfined system \cite{cas90}.
This implies that the confinement increases the effectiveness of the friction mechanism:
Fast particles, for which the friction force tends to zero (see Eq.~\eqref{langevin}), eventually decelerate due to the harmonic restoring force and re-enter the momentum range where the friction is sizable, thus increasing overall dissipation.
The energy PDF \eqref{stationary-density} is the leading order of an expansion in terms of $\Omega^{-1}$ \cite{bor87}.
More precisely, we have,
\begin{align}
P(E,\alpha) = \frac{P_E(E)}{2 \pi} \bigg[1 + \frac{f_1(E,\alpha)}{\Omega} + \mathcal{O}(\Omega^{-2}) \bigg], \label{underdamped-expansion}
\end{align}
which allows us to find corrections for finite frequencies.
These correction terms will depend on the angle $\alpha$ and thus violate energy equipartition.
The explicit function $f_1(E,\alpha)$ is derived in the SM, by plugging the expansion \eqref{underdamped-expansion} into Eq.~\eqref{klein-kramers-energy} and equating terms of the same order in $\Omega$.
The PSPDF corresponding to this first-order result is shown in Fig.~\ref{fig:underdamped-joint-dist}.
Except at low energies, this clearly exhibits deviations from equiprobable energy surfaces, which would appear as straight lines, and shows excellent agreement between our theory and simulation results.


\textbf{Large energies $E \gg 1$.} 
Importantly, the asymptotic behavior $P_E(E) \propto E^{-1/D}$ \eqref{stationary-density} is not only valid for large $\Omega$, but is the generic case for large energies.
Physically, this is due to the nonlinear behavior of the friction force, which tends to zero for large momenta and thus leads to underdamped behavior at large energies.
The large energy tails of the distribution are thus universally described by the underdamped approximation.
This can be understood by noting that, except in a small \enquote{strip} around $\alpha = 0$ or $\alpha = \pi$, where $E \sin^2(\alpha)$ is small, the operator $\mathcal{L}_E$ in Eq.~\eqref{klein-kramers-energy} is of order $E^{-1}$.
Similar to Eq.~\eqref{underdamped-expansion}, we can then expand the PSPDF,
\begin{align}
P(E,\alpha) \simeq \frac{P_E(E)}{2\pi} \bigg[ 1 + \frac{g_1(\alpha)}{E} + \mathcal{O}(E^{-2}) \bigg] . \label{energy-expansion-outside}
\end{align}
Expanding $\mathcal{L}_E$ for large energies, we can then determine the function $g_1(\alpha)$ (see SM).
The resulting asymptotic PSPDF up to order $E^{-1}$ reads,
\begin{align}
P(E,&\alpha) \simeq N E^{-\frac{1}{D}} \bigg[1 - \frac{\sqrt{2}}{D}  E^{-1/2} + \frac{1}{D^2} E^{-1} \nonumber \\
& +  \Big[ \frac{1}{2 \Omega} \Big( \big(1 + \frac{1}{D}\big) \sin(2\alpha) - \cot(2 \alpha) \Big) \Big] E^{-1} \bigg], \label{large-energy}
\end{align}
where $N$ is a normalization constant.
The first three terms on the right hand side stem from the large energy expansion of $P_E(E)$, whereas the remaining term is the angle-dependent correction $g_1(\alpha)$.
Due to the cotangent appearing in the latter, Eq.~\eqref{large-energy} diverges at $\alpha = 0$ and $\alpha = \pm \pi$.
These singularities occur because in Eq.~\eqref{large-energy}, we expanded for large $E$, assuming that $2 E \sin^2(\alpha)$ is large, which breaks down close to $\alpha = 0$ and $\alpha = \pm \pi$.
Inside the strip, we express Eq.~\eqref{klein-kramers-energy} as a function of $z = \sqrt{2 E} \alpha$ and $E$ and again expand for large energies.
Asymptotic matching of both expansions and the condition that $P(E,\alpha)$ is a periodic function of $\alpha$ fixes any occurring integration constants.
We will detail this procedure in a longer publication \cite{upcoming}, the result up to order $E^{-1}$ is given in Eq.~(S12) in the SM.
The main conclusion from Eq.~\eqref{energy-expansion-outside} is that, for large energies, the PSPDF decays as a power law in energy, with small angle-dependent corrections.

\begin{figure}
\includegraphics[width=0.48\textwidth, clip, trim=0cm 0cm 0cm 0cm]{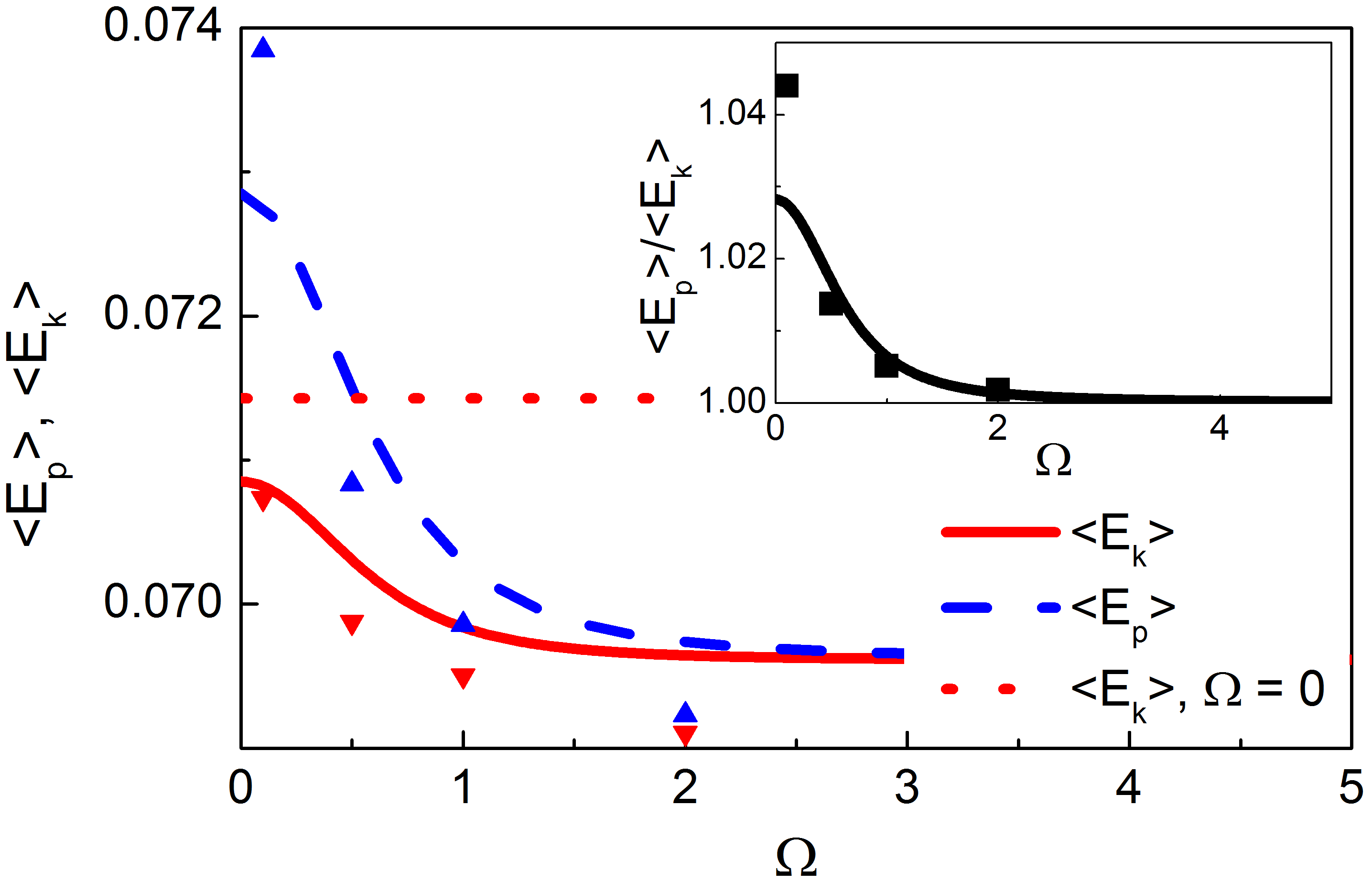}
\caption{Average potential and kinetic energy for $D = 0.1$ from the third-order small-$D$ expansion as a function of trap frequency. The kinetic energy of a free particle without the potential is shown for comparison. \textbf{Inset}: Ratio of potential and kinetic energy. For large frequencies, the energy is equally distributed between the potential and kinetic degree of freedom. For small frequencies, the potential energy is larger than the kinetic one. The symbols are the results of numerical simulations, deviations from the theoretical curves are due to the truncation of the expansion; better agreement but overall smaller effects are found for smaller values of $D$.}
\label{fig-equipartition}
\end{figure}

\textbf{Deep lattices $D \ll 1$.}
For small $D$, the atoms are typically slow and thus the friction is approximately linear.
While the large energy tails are still given by Eq.~\eqref{energy-expansion-outside}, the power law decays very fast for small $D$ and the center part dominates the statistics of the system.
In this regime, we thus expect the BG equilibrium distribution Eq.~\eqref{boltzmann-potential} to approximate the PSPDF.
Indeed, by rescaling $\tilde{x} = x/\sqrt{D}$ and $\tilde{p}=p/\sqrt{D}$ in Eq.~\eqref{klein-kramers}, we see that in the limit $D \rightarrow 0$, we recover the usual Stokes friction result and thus the BG distribution \eqref{boltzmann-potential}.
We define an auxiliary function $h({\tilde{x}},{\tilde{p}})$ via,
\begin{align}
P({\tilde{x}},{\tilde{p}}) = \underbrace{\frac{1}{2 \pi} e^{-\frac{{{\tilde{p}}^2+\tilde{x}}^2}{2}}}_{P_\text{BG}(\tilde{x},\tilde{p})} h({\tilde{x}},{\tilde{p}}) ,
\end{align}
and expand the latter with respect to $D$,
\begin{align}
h({\tilde{x}},{\tilde{p}}) \simeq 1 + D h_1({\tilde{x}},{\tilde{p}}) + D^2 h_2({\tilde{x}},{\tilde{p}}) + \mathcal{O}(D^3) \label{small-d-expansion}
\end{align}
Plugging this into Eq.~\eqref{klein-kramers} and equating coefficients, we find a recursive set of equations for $h_1({\tilde{x}},{\tilde{p}})$, $h_2({\tilde{x}},{\tilde{p}})$ and so on.
As it turns out, the functions $h_{n}({\tilde{x}},{\tilde{p}})$ are polynomials of up to order $4 n$ in ${\tilde{x}}$ and ${\tilde{p}}$, which reduces the problem to solving linear equations for the coefficients.
In order for the expansion \eqref{small-d-expansion} to be valid, $D \tilde{x}^k \tilde{p}^l$ with $k+l=4$ has to be small.
The above expansion thus accurately describes the center part of the PSPDF for small $D$.
Contrary to the underdamped approximation, this small noise, small energy expansion places no restrictions on $\Omega$, allowing us to explore the overdamped regime $\Omega \ll 1$.
To first order, the resulting PSPDF is,
\vspace*{-2em}
\begin{widetext}
\begin{align}
P(\tilde{x},\tilde{p}) = \frac{e^{-\frac{\tilde{p}^2+\tilde{x}^2}{2}}}{2 \pi} \bigg[ 1 + \frac{D}{4 (3+4\Omega^2)} \Big[ 3 \tilde{p}^4 + 18 \tilde{x}^2 - 27 + \Big( 4 \tilde{p}^3 \tilde{x} - 12 \tilde{p} \ \tilde{x} \Big) \Omega + \Big(3(\tilde{p}^2+\tilde{x}^2)^2 - 24 \Big) \Omega^2 \Big] \bigg] . \label{small-energy}
\end{align}
\end{widetext}
In practice, we perform the expansion up to order $D^3$; the resulting expression agrees with simulations of the PSPDF and the small-energy behavior of the marginal energy distribution $P_E(E)$ (see Fig.~\ref{fig:energy-dist}).
The expansion procedure is detailed in the SM.
The PSPDF has a number of interesting features.
For large frequencies $\Omega \gg 1$, the last term on the r.h.s.~of Eq.~\eqref{small-energy} dominates and it can be expressed as a function of the Hamiltonian $H(x,p) = (p^2+x^2)/2$, which corresponds to the underdamped limit Eq.~\eqref{stationary-density}.
For small frequencies $\Omega \ll 1$, this is no longer the case and equipartition breaks down.
Instead, we find that the average potential energy is larger than the kinetic one, while both the kinetic and potential energy and thus the total energy of the system increase as $\Omega \rightarrow 0$, see Fig.~\ref{fig-equipartition}.
This is in stark contrast to the case of linear friction, where the energy of the system is always equi-distributed between kinetic and potential energy and the total energy is independent of the frequency.
Even for very low frequencies, the average kinetic energy is reduced (compared to the free particle case, dashed line in Fig.~\ref{fig-equipartition}) by introducing the confining potential, supporting the notion that confinement increases the effectiveness of the friction mechanism.
The breakdown of equipartition means that the temperature of the system is not uniquely defined, as we necessarily have different effective temperatures governing the kinetic and potential degrees of freedom.

\textbf{Discussion.}
We have investigated the statistical mechanics of cold atoms subject to Sisyphus cooling and a harmonic confinement in different regimes.
Let us come back to the two questions we posed in the beginning:
Is the PSPDF a function of the Hamiltonian?
In the three limiting cases discussed above -- namely for large frequencies, large energies and small $D$ -- the answer is yes, at least to leading order.
The underdamped limit, which for generic systems requires large frequencies, also describes the large energy behavior for arbitrary frequencies.
The physical reason for this is that dissipation is weak for fast particles due to the nonlinear friction force.
However, as we have shown, there are corrections to the leading order results, which cannot be expressed as a function of the Hamiltonian. 
Generally, contrary to BG statistics, equal energy thus does not imply equal probability.
In particular, for moderate values of $D$ and small frequencies, there may be sizable deviations from energy equipartition \footnote{From numerical simulations, we find e.g.~$\protect\langle E_p \protect\rangle \approx  1.2 \protect\langle E_k \protect\rangle$ for $D = 1/3$ and $\Omega = 0.1$.}.
Is the PSPDF exponential in the Hamiltonian?
Here, the answer is affirmative only for deep lattices $D \rightarrow 0$, where the PSPDF reduces to the BG result.
For finite $D$, however, the tails of the PSPDF will be power-laws, $P(x,p) \sim (H(x,p))^{-1/D}$, leading to qualitative differences from BG statistics.

From an experimental point of view, the deviations from equipartition mean that care needs to be taken when assigning a temperature to the particle cloud, as the latter -- if interpreted in terms of kinetic energy -- is generally not equal to the atoms' potential energy.
The confinement lowers the kinetic temperature of the atoms, potentially improving the cooling mechanism 
\footnote{Within the semiclassical approximation used here, the confinement enhances the cooling mechanism. However, there are other effects, e.g.~Stark shifts induced by the confining field, that need to be taken into account \cite{upcoming}}.
In a recent experiment \cite{sag12}, the atoms were equilibrated with the lattice in a dipole trap, before being released in order to measure their superdiffusive motion.
The trapping phase corresponds precisely to the situation discussed in this Letter, with $\Omega \ll 1$ and $1/5 \lesssim D \lesssim 1/3$.
On the one hand, this demonstrates that the parameter regime where the deviations from BG and equipartition are relevant is accessible in experiment.
On the other hand, as we have shown, the steady state energy PDF in the trap has power-law tails, which may change the diffusive dynamics itself \cite{hir11,hir12}, compared to the narrow initial condition assumed in the theoretical discussion of the unconfined system \cite{dec14,bar14}.
In summary, our analysis opens up an exciting pathway to an anomalous equilibrium statistical mechanics, implemented in an existing experimental system.

\textbf{Acknowledgments.} This work was supported by the Israel Science Foundation.

\bibliography{bib}

\pagebreak

\appendix

\part*{Supplementary Material}

\renewcommand{\theequation}{S\arabic{equation}}
\renewcommand{\figurename}{FIG. S}

\setcounter{equation}{0}

\section*{Underdamped approximation}
Starting out from Eq.~(9) in the main text,
\begin{align}
P(E,\alpha) = \frac{P_E(E)}{2 \pi} \bigg[1 + \frac{f_1(E,\alpha)}{\Omega} + \mathcal{O}(\Omega^{-2}) \bigg], \label{large-omega-expansion}
\end{align}
and plugging this into the stationary Kramers equation (4),
\begin{align}
\bigg[\partial_{\alpha} + &\Omega^{-1} \mathcal{L}_E(\alpha) \bigg] P(E,\alpha) = 0  \label{klein-kramers-energy-supp}
\end{align}
with
\begin{align} 
\mathcal{L}_E(\alpha) &= \partial_\alpha \frac{\sin(\alpha)\cos(\alpha)}{1+2 E \sin^2(\alpha)} + \partial_E \frac{2 E \sin^2(\alpha)}{1+2 E \sin^2(\alpha)}  \nonumber \\
& + D \bigg[ 2 \partial_\alpha \sin(\alpha) \cos(\alpha) \partial_E  + \frac{1}{2 E} \partial_\alpha \cos^2(\alpha) \partial_\alpha \nonumber \\
& + (\sin^2(\alpha) - \cos^2(\alpha)) \partial_E + 2 \sin^2(\alpha) \partial_E E \partial_E \bigg] , \nonumber
\end{align}
we find to order $\Omega^{-1}$,
\begin{align}
P_E(E) \partial_{\alpha} f_1(E,\alpha) + \mathcal{L}_E(\alpha) P_E(E) = 0 .
\end{align}
Solving for $f_1(E,\alpha)$, we have,
\begin{align}
f_1(E,\alpha) = P_E^{-1}(E) \int_{0}^{\alpha} \text{d}\alpha' \ \mathcal{L}_E(\alpha') P_E(E) + f_{1,0}(E), \label{large-omega-correction}
\end{align}
where the integration constant $f_{1,0}(E)$ is some function of energy.
Performing the integral in Eq.~\eqref{large-omega-correction}, we obtain for $f_1(E,\alpha)$,
\begin{widetext}
\begin{align}
f_1(E,\alpha) = &-\frac{\arctan^{*}\big(\sqrt{1+2 E} \tan(\alpha) \big)}{\sqrt{1+2 E}} \bigg[ \frac{1}{1 + 2 E} + \frac{\partial_E P_E(E)}{P_E(E)} \bigg] + \frac{E}{1+2E} \frac{\sin(2 \alpha)}{E \cos(2\alpha) - E - 1} \nonumber \\
& \qquad - \bigg[ \frac{D}{2} \sin(2 \alpha) + \alpha \bigg] \frac{\partial_E P_E(E)}{P_E(E)} - \Big(D \alpha - \frac{1}{2}\sin(2\alpha) \Big) \frac{\partial_E E \partial_E P_E(E)}{P_E(E)} + f_{1,0}(E),
\end{align}
where we define,
\begin{align}
\arctan^{*}(x \tan(\alpha)) = \arctan(x \tan(\alpha)) + k \pi \quad \text{for} \quad \frac{(2 k - 1) \pi}{2} < \alpha < \frac{(2 k + 1) \pi}{2} .
\end{align}
\end{widetext}
Since the system is symmetric with respect to inversion $(x,p) \rightarrow (-x,-p)$, the function $f_1(E,\alpha)$ has to be $\pi$-periodic, $f_1(E,\alpha+\pi)-f_1(E,\alpha)=0$.
After some algebra, we get a condition on $P_E(E)$, which is precisely Eq.~(6).
The resulting function $f_1(E,\alpha)$ is shown in Fig.~S 1.
\begin{figure}
\includegraphics[width=0.48\textwidth]{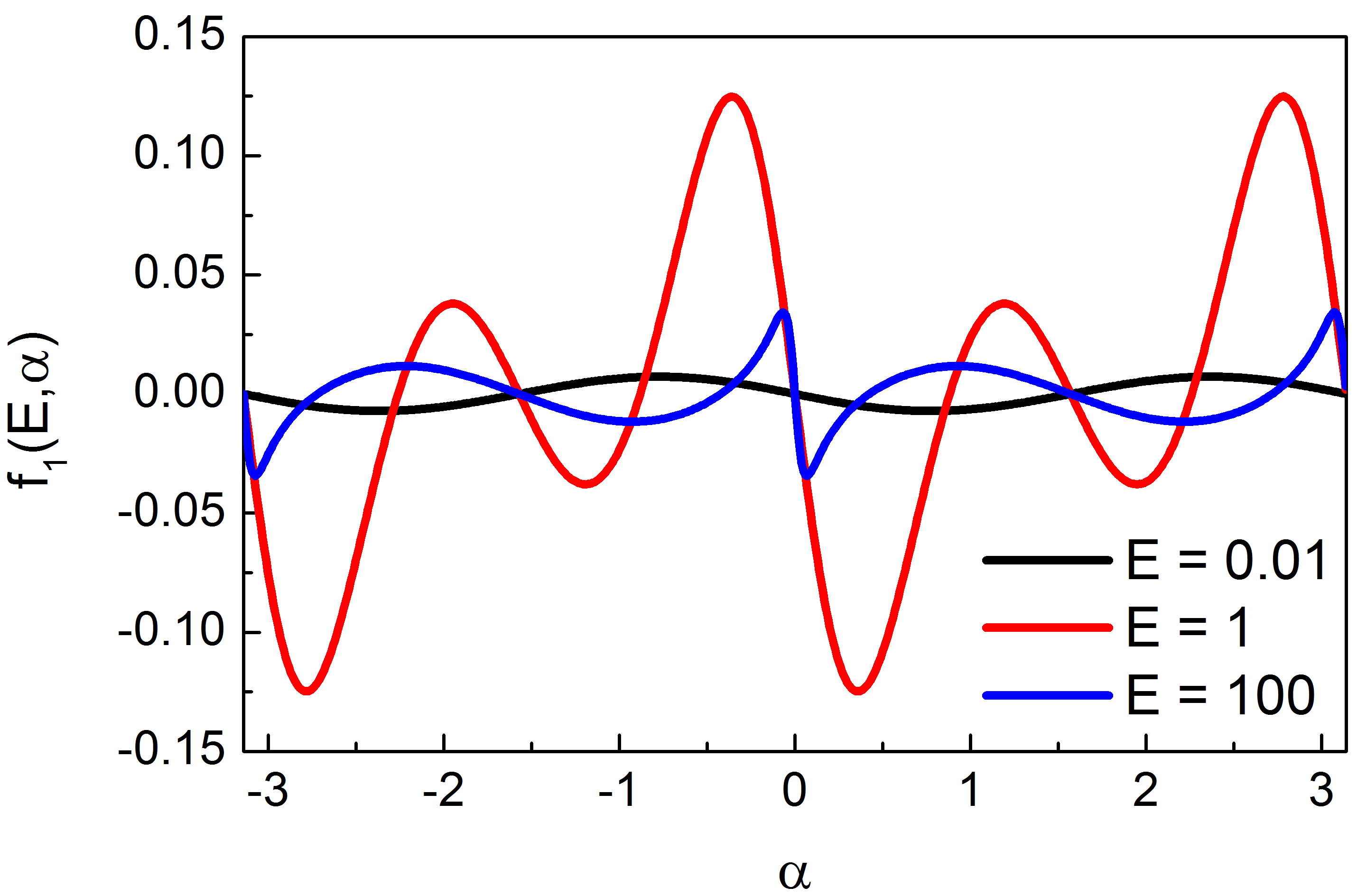}
\caption{The first-order angle-dependent correction term $f_1(E,\alpha)$ to the phase-space distribution as a function of $\alpha$ for three different values of the energy and $D = 1/3$.}
\label{fig:anglefunction}
\end{figure}
Note that it is small for both small and large energies, where the phase-space density depends only weakly on the angle.
For intermediate energies, on the other hand, the corrections may be sizable.
In the above we so far ignored the integration constant $f_{1,0}(E)$.
To compute the latter, we need to evaluate the order $\Omega^{-2}$ correction $f_2(E,\alpha)$,
\begin{align}
f_2(E,\alpha) &= P_E^{-1}(E) \int_{0}^{\alpha} \text{d}\alpha' \ \mathcal{L}_E(\alpha') f_1(E,\alpha') P_E(E) \nonumber \\
& \qquad + f_{2,0}(E).
\end{align}
Demanding the $\pi$-periodicity of $f_2(E,\alpha)$ then provides us with a condition on $f_{1,0}(E)$.
Using that $f_1(E,\alpha)$ is antisymmetric with respect to $\alpha = \pi/2$ and the symmetry of $\mathcal{L}_E(\alpha)$, the condition on $f_{1,0}(E)$ turns out to be,
\begin{align}
\bigg[\partial_E \Big(1-\frac{1}{\sqrt{1+2E}}\Big) + D \partial_E E \partial_E \bigg] f_{1,0}(E) P_E(E) = 0. 
\end{align}
Since $P_E(E)$ by itself satisfies this equation, $f_{1,0}(E)$ can be an arbitrary constant.
However, the normalization of the total PPDF fixes $f_{1,0}(E) = 0$.

\section*{Large-energy expansion}
The procedure employed to find the function $g_1(\alpha)$ in the large-energy expansion,
\begin{align}
P(E,\alpha) = \frac{P_E(E)}{2 \pi} \bigg[1 + \frac{g_1(\alpha)}{E} + \mathcal{O}(E^{-2}) \bigg], \label{large-energy-expansion}
\end{align}
is similar to the large-frequency expansion Eq.~\eqref{large-omega-expansion}, as the behavior for large energies is underdamped as well.
In order to be consistent in the expansion, we plug the above into Eq.~\eqref{klein-kramers-energy-supp} and expand the resulting equation for large $E$.
To order $E^{-1}$, we then find,
\begin{align}
\partial_{\alpha} g_1(\alpha) = \frac{2 D \cos^2(\alpha) + (2 + D)\cos(2\alpha) + D \cot(\alpha)^2 }{2 D \Omega}.
\end{align}
Integrating over $\alpha$, we get,
\begin{align}
g_1(\alpha) = \frac{1}{2 \Omega} \Big( \big(1 + \frac{1}{D}\big) \sin(2\alpha) - \cot(2 \alpha) \Big) + g_{1,0}, \label{large-energy-correction}
\end{align}
where $g_{1,0}$ is a constant independent of $E$ and $\alpha$.
As was the case with the large-frequency expansion above, $g_{1,0}$ has to be determined from the second order by demanding periodicity.
However, this is not straightforward.
The first order correction Eq.~\eqref{large-energy-correction} diverges at $\alpha = 0$ and $\alpha = \pm \pi$, the same is true for the second order.
The latter is thus not valid in the entirety of phase-space, so we cannot impose periodicity, which is a global property.
The divergence is due to the presence of the strip discussed in the main text, where the momentum is of order 1.
We have to take into account the contribution from the strip to impose periodicity and find $g_{1,0}=0$, which agrees with $f_{1,0}(E) = 0$ from the large-frequency result.
The solution for $P(E,\alpha)$ outside and inside the strip and up to order $E^{-1}$ reads [20],
\begin{widetext}
\begin{align}
P(E,\alpha) &\simeq N E^{-\frac{1}{D}} \nonumber \\
&\times  \left\lbrace \begin{array}{ll}
1 - \frac{\sqrt{2}}{D}  E^{-1/2} + \Big[ \frac{1}{D^2} + \frac{1}{2 \Omega} \Big( \big(1 + \frac{1}{D}\big) \sin(2\alpha) - \cot(2 \alpha) \Big) \Big] E^{-1},& \text{for} \quad 2 E \sin^2(\alpha) \gg 1 \\[2 ex]
1 - \Big[\frac{\sqrt{2}}{D} + \frac{\sqrt{E} \alpha}{ \Omega (1+2 E \alpha^2)} \Big] E^{-1/2} + \Big[\frac{1}{D^2} + \frac{1}{2 \Omega^2} \Big[ \frac{2 D -1}{(1+2 E \alpha^2)^2} + \frac{D-D^2 + 2 \sqrt{2 E} \alpha \Omega}{D(1+2 E \alpha^2)} \Big] \Big] E^{-1},  & \text{for} \quad 2 E \sin^2(\alpha) \lesssim 1 .
\end{array} \right. \label{large-energy-solution}
\end{align}
\end{widetext}

\section*{Small-$D$ expansion}
The stationary Klein-Kramers equation (3) in terms of the rescaled variables $\tilde{x}$ and $\tilde{p}$ reads,
\begin{align}
\Bigg[\Omega \bigg[ - \tilde{p} \partial_{\tilde{z}} + {\tilde{z}} \partial_{\tilde{p}} \bigg] + \partial_{\tilde{p}} \bigg[ \frac{{\tilde{p}}}{1+D {\tilde{p}}^2} + \partial_{\tilde{p}} \bigg] \Bigg] P({\tilde{z}},{\tilde{p}}) = 0 .
\end{align}
It is immediately apparent that this reduces to the equation for linear friction in the limit $D \rightarrow 0$.
Introducing the auxiliary function $h(\tilde{x},\tilde{p})$ as in Eq.~(14), we get an equation for the latter,
\begin{align}
\Big[\mathcal{L}_0 &+ D \mathcal{L}_1 + D^2 \mathcal{L}_2\Big] h(\tilde{x},\tilde{p}) = 0 \\
\mathcal{L}_0 &= \Omega(\tilde{x} \partial_{\tilde{p}} - \tilde{p} \partial_{\tilde{x}}) - \tilde{p} \partial_{\tilde{p}} + \partial_{\tilde{p}}^2 \nonumber \\
\mathcal{L}_1 &= 2 \Omega (\tilde{x} \tilde{p}^2 \partial_{\tilde{p}} - \tilde{p}^3 \partial_{\tilde{x}}) - 3 \tilde{p}^2 + \tilde{p}^4 - 3 \tilde{p}^3 \partial_{\tilde{p}} + 2 \tilde{p}^2 \partial_{\tilde{p}}^2 \nonumber \\
\mathcal{L}_3 &= \Omega (\tilde{x} \tilde{p}^4 \partial_{\tilde{p}} - \tilde{p}^5 \partial_{\tilde{x}}) - \tilde{p}^4 + \tilde{p}^6 - 2 \tilde{p}^5 \partial_{\tilde{p}} + \tilde{p}^4 \partial_{\tilde{p}}^2 \nonumber .
\end{align}
Expanding $h(\tilde{x},\tilde{p})$ for small $D$ (see Eq.~(15)), we find to order $D^{0}$,
\begin{align}
\mathcal{L}_0 h_0(\tilde{x},\tilde{p}) = 0,
\end{align}
which is solved by $h_0(\tilde{x},\tilde{p}) = \frac{1}{2 \pi}$, leading to the normalized BG distribution.
To order $D^{1}$, we then have,
\begin{align}
\mathcal{L}_0 h_1(\tilde{x},\tilde{p}) + \mathcal{L}_1 \frac{1}{2 \pi} = \mathcal{L}_0 h_1(\tilde{x},\tilde{p}) + \frac{1}{2 \pi} (\tilde{p}^4-3\tilde{p}^2) = 0 . \label{small-d-first-order}
\end{align}
Now if $h_1(\tilde{x},\tilde{p})$ is assumed to be of polynomial form,
\begin{align}
h_1(\tilde{x},\tilde{p}) = \sum_{k=0}^{n_1} \sum_{l=0}^{n_0-k} a_{1;k l} \tilde{x}^{k} \tilde{p}^{l} \label{small-d-g1} ,
\end{align}
then, since $\mathcal{L}_1$ at most reduces the order of the polynomial by $2$, we need to have $n_1 = 4$ to satisfy Eq.~\eqref{small-d-first-order}.
Indeed, plugging in Eq.~\eqref{small-d-g1}, we find a closed set of equations for the coefficients $a_{1;k l}$, $a_{1,0 0}$ being fixed by normalization.
Similarly, we find that the $n$-th order function $h_{n}(\tilde{x},\tilde{p})$ is a polynomial of order $4 n$ in $\tilde{x}$ and $\tilde{p}$,
\begin{align}
h_n(\tilde{x},\tilde{p}) = \sum_{k=0}^{4 n} \sum_{l=0}^{4 n-k} a_{n;k l} \tilde{x}^{k} \tilde{p}^{l} .
\end{align}

\end{document}